\documentclass[aps,10pt,prd,twocolumn,preprintnumbers,amsmath,amssymb,nofootinbib,superscriptaddress,a4paper,showpacs]{revtex4-1}

\usepackage[breaklinks]{hyperref}
\usepackage[latin1]{inputenc}
\usepackage{graphicx}
\usepackage{color}
\usepackage{amsmath,amssymb,amsfonts,amsthm}
\usepackage{bm,bbm}
\usepackage{soul}

\newcommand{\be}{\begin{equation}}
\newcommand{\ee}{\end{equation}}
\newcommand{\ds}{d_{\rm S}}
\newcommand{\s}{\sigma}
\newcommand{\g}{\gamma}
\newcommand{\la}{\lambda}
\newcommand{\Eqref}[1]{Eq.~(\ref{#1})}

\def\cS{{\cal S}}

\def\cP{{\cal P}}

\def\rmd{d}
\newcommand{\p}{\partial}
\newcommand{\oarX}[1]{\href{http://arxiv.org/abs/#1}{{\ttfamily\com #1}}}
\newcommand{\arX}[1]{\href{http://arxiv.org/abs/#1}{{\ttfamily\com arXiv:#1}}}
\newcommand{\doin}[6]{\href{http://dx.doi.org/#1}{\cob #2 #3 {\bf #4}, #5 (#6)}}
\newcommand{\ndoin}[6]{\href{#1}{\cob #2 #3 {\bf #4}, #5 (#6)}}
\newcommand{\doij}[5]{\href{http://dx.doi.org/#1}{\cob #2 #3 (#5) #4}}
\newcommand{\tia}[1]{}
\def\com{\color{magenta}}
\def\cob{\color{blue}}

\begin{document}

\title{Probing the quantum nature of spacetime by diffusion}

\author{Gianluca Calcagni}
\email{calcagni@iem.cfmac.csic.es}
\affiliation{Instituto de Estructura de la Materia, CSIC, Serrano 121, 28006 Madrid, Spain}
\author{Astrid Eichhorn}
\email{aeichhorn@perimeterinstitute.ca}
\affiliation{Perimeter Institute for Theoretical Physics, 31 Caroline Street N, Waterloo, N2L 2Y5, Ontario, Canada}
\author{Frank Saueressig}
\email{F.Saueressig@science.ru.nl}
\affiliation{Institute for Mathematics, Astrophysics and Particle Physics, Radboud University Nijmegen, Heyendaalseweg 135, 6525 AJ Nijmegen, The Netherlands}

\begin{abstract} 
Many approaches to quantum gravity have resorted to diffusion processes to characterize the spectral properties of the resulting quantum spacetimes. We critically discuss these quantum-improved diffusion equations and point out that a crucial property, namely positivity of their solutions, is not preserved automatically. We then construct a novel set of diffusion equations with positive semidefinite probability densities, applicable to asymptotically safe gravity, Ho\v{r}ava-Lifshitz gravity and loop quantum gravity. These recover all previous results on the spectral dimension and shed further light on the structure of the quantum spacetimes by assessing the underlying stochastic processes. Pointing out that manifestly different diffusion processes lead to the same spectral dimension, we propose the probability distribution of the diffusion process as a refined probe of quantum spacetime.
\end{abstract}

\date{April 26, 2013}



\preprint{\doin{10.1103/PhysRevD.87.124028}{Phys.\ Rev.}{D}{87}{124028}{2013} [\arX{1304.7247}]}

\maketitle

\section{Introduction}

In quantum gravity it is commonly expected that quantum fluctuations will lead to a spacetime structure that departs from a smooth classical manifold at short scales. A common theme shared by many quantum gravity candidates focuses on the characterization of the resulting quantum spacetimes and possible observable consequences. In this context, the analysis of the spectral properties, which can be probed by the diffusion of a fiducial test particle on the effective quantum spacetime, have turned out to be quite useful. In particular, they allow discriminating between the pictures of quantum spacetime emerging from different theories.

The simplest spectral quantity carrying nontrivial information of the spacetime structure is the spectral dimension $\ds$. This notion of dimensionality, which is distinct from the topological or Hausdorff dimension, arises in the context of a diffusion process: From the probability density $P(x,x',\sigma)$ of a diffusing particle on a background, one can define a return probability $\cP(\sigma) = V^{-1}\int_x P(x,x,\sigma)$. Herein, $x$ and $x'$ denote coordinates on the Euclidean spacetime, $V$ is the volume and $\sigma$ an external diffusion time. The spectral dimension for a background with fixed dimensionality is then defined as\footnote{ At this stage the derivative symbols are not strictly necessary due to the presence of the limit, but will become important in the more general case of a scale-dependent spectral dimension.}
\be\label{specdef}
\ds = -2\, \underset {\sigma \rightarrow 0}{ \lim} \frac{\p\ln \cP(\sigma)}{\p \ln \sigma} \,.
\ee

By now, the spectral dimension has been determined within several quantum gravity approaches,
including causal dynamical triangulations (CDT) in four \cite{Ambjorn:2005db,Gorlich:2011ga} and three spacetime
dimensions \cite{Benedetti:2009ge,Anderson:2011bj}, Euclidean dynamical triangulations (EDT) \cite{Laiho:2011ya},
loop quantum gravity (LQG) \cite{Modesto:2008jz}, asymptotically safe gravity (or quantum Einstein gravity, QEG) \cite{Lauscher:2005qz,Lauscher:2005xz,Reuter:2011ah,Rechenberger:2012pm}, nonlocal  gravity \cite{Mod11},
and also Ho\v{r}ava-Lifshitz (HL) gravity \cite{Horava:2009if,Sotiriou:2011mu}. Remarkably, many of these works
assess that $\ds = 2$ at microscopic scales \cite{Car09,fra1,Car10}.
Subsequently, it actually turned out to be useful to generalize the definition \eqref{specdef} by dropping the
limit $\sigma \rightarrow 0$ and considering $\ds(\sigma)$ as a function of the diffusion time. This generalization effectively allows one to characterize a spacetime with multifractal structures
where the spectral properties change on various typical length scales, e.g., when considering the spacetime structure at classical, Planckian, and sub-Planckian distances.

To arrive at the return probability, two different routes have been followed in the literature. The Monte Carlo approaches to quantum gravity, foremost CDT and EDT, approximate the quantum spacetime by a piecewise linear manifold built from $d$-dimensional simplices \cite{Ambjorn:2012jv}. In this setting, the quantum nature of spacetime is encoded in the gluing of the building blocks. The return probability is then obtained by studying a standard random walk on the resulting discrete quantum spacetime. In other words, a classical random walk is used to capture the fractal features of the geometry. The probe particle thereby has equal probabilities for moving from its current simplex to one of its neighbors in each time step \cite{Ambjorn:2005db,Benedetti:2009ge,Laiho:2011ya,Anderson:2011bj}. This construction implies that, after each time step, the probe particle has to be located somewhere on the geometry, guaranteeing a positive semidefinite probability density $P(x, x^\prime, \sigma)$.

Within analytical approaches to quantum gravity, it has been proposed that quantum effects in spacetime, foremost a dynamical dimensional change, can be captured by a modification of the Laplacian operator appearing in the classical diffusion equation \cite{Lauscher:2005qz,Lauscher:2005xz,Reuter:2011ah,Rechenberger:2012pm,Modesto:2008jz,Horava:2009if,Sotiriou:2011mu,frc4, Mod11}.  We will call this new operator ``generalized Laplacian'' or, as done in probability theory, spatial generator. Thus, one uses an anomalous diffusion equation on classical flat spacetime to mimic the quantum structure. The underlying physical picture is that the effective metric ``seen'' by the diffusing particle actually depends on the momentum of the probe.\footnote{Although a scale-dependent metric does not seem to make sense at a first glance, one should note that a unique, dynamically determined metric emerges as the solution to the full quantum equations of motion in the infrared. Determining momentum scales with respect to this metric then gives meaning to the notion of a scale-dependent family of metrics.} Expressing this metric through a fixed reference scale leads to a modified diffusion equation providing an effective description of  the propagation of the probe particle on the quantum gravity background. For asymptotically safe gravity \cite{Reuter:2012id,Reuter:1996cp,Codello:2008vh, Benedetti:2009rx} in $d$ dimensions, this procedure gives rise to a multifractal structure where the spectral dimension is of the form \cite{Reuter:2011ah}
\be
\ds = \frac{2d}{2+\delta} \, .
\ee
Here, the parameter $\delta$ captures the quantum effects and actually depends on the probed length scale. One can identify three characteristic regimes where the spectral dimension is approximately constant over many orders of magnitude. At large distances, one encounters the classical regime where $\delta = 0$ and the spectral dimension agrees with both the Hausdorff and topological dimension of the spacetime. At smaller distances one first encounters a semiclassical regime with $\delta = d$, before entering into the fixed-point regime with $\delta = 2$. 

 Let us clarify some conceptual points arising in this approach of characterizing the quantum spacetime. Crucially, diffusion processes probe the properties of the effective background spacetime only. There is no relation to fluctuations of the background which would be encoded in the graviton propagator. Accordingly, in our context the spectral dimension is probably not directly understood as ``the dimension of momentum space,'' as conjectured in fractal geometry \cite{Akk2,Akk12}. Furthermore, the diffusing particle is a fictitious probe, and no backreaction of the particle on the background is included. Finally, the diffusion time $\sigma$ is an external time, thus it is not related to a (causal) propagation of a matter field on the quantum spacetime. In other words, the scaling of the graviton and also matter propagators in quantum gravity need not show an effective two-dimensionality, even in the case $\ds =2$ \cite{Eichhorn:2010tb,Groh:2010ta,Vacca:2010mj,Rosten:2011mf,Eichhorn:2012va}. We thus stress that the spectral dimension is a useful tool to encode properties of an effective spacetime in quantum gravity, but it is not necessarily related to the propagation of a physical particle.

A potential issue related to the use of the modified diffusion equations proposed in the quantum gravity literature \cite{Lauscher:2005qz, Reuter:2011ah,Rechenberger:2012pm,Modesto:2008jz,Horava:2009if,Sotiriou:2011mu}
is their potential risk of giving rise to ``negative probabilities.'' In fact, we will explicitly demonstrate in the next section that the solutions of these modified equations are not positive semidefinite, so that the interpretation as probabilistic processes is lost. We take this observation as a clue that these descriptions are incomplete, and formulate several proposals for restoring the probabilistic nature of the modified diffusion processes in Secs.\ \ref{sect.3} to \ref{sect.6}. This allows us to make another step forward in the exploration of quantum spacetimes: The diffusion probability encodes more information on the underlying spacetime than the spectral dimension alone, and permits one to study the quantum properties of spacetime in more detail. In particular, we provide a setup which allows one to construct diffusion equations accommodating generic nontrivial scaling properties even within an anisotropy between space and time. We thus expect our setup to be relevant to several quantum gravity approaches, encompassing asymptotic safety, Ho\v{r}ava-Lifshitz gravity and loop quantum gravity, as well as the tentative continuum limit of CDT and EDT.

\section{Diffusion equations with no probabilistic interpretation}
\label{negprob}
\noindent
Nonrelativistic Brownian motion in $d$ dimensions with initial condition $P(x,x',0)= \delta(x-x')$ is described by
\be\label{BM}
\left(\partial_{\sigma} - \nabla_x^2\right) P(x,x',\sigma)=0.
\ee
One way to encode quantum-gravity effects is to modify the Laplacian $\nabla^2 = g^{\mu \nu}\nabla_{\mu} \nabla_{\nu}$
by replacing the classical spacetime metric $g^{\mu \nu}$ by a suitable quantum object \cite{Lauscher:2005qz},
\be\label{absdiff}
\left(\partial_{\sigma} - \langle \Delta \rangle \right) P(x,x',\sigma)=0 \, , 
\ee
where $\langle \Delta \rangle$ is a generalized Laplacian. In asymptotically safe gravity, this object is naturally constructed from the vacuum expectation value of the quantum metric which can be computed from the scale-dependent 
flowing action $\Gamma_k$. Similarly, heuristic arguments in loop quantum gravity allow one to obtain a quantum version of $g_{\mu \nu}$ based on the area operator \cite{Modesto:2008jz}, while anisotropic Ho\v{r}ava-Lifshitz gravity resorts to a Laplace operator that contains higher powers of spatial derivatives \cite{Horava:2009if,Sotiriou:2011mu}. 

The solution $P$ for these cases can be evaluated
 by resorting to a flat reference spacetime at macroscopic scales.
 One can then express $P(x,x',\sigma)$ through its Fourier transform. For example, the application of renormalization group (RG) improvement schemes within asymptotic safety yields a solution of the form
\cite{Lauscher:2005qz,Reuter:2011ah}
\be\label{PQEG}
P_{\textsc{qeg}}(x,x',\sigma) = \int \frac{d^dp}{(2\pi)^d} \, e^{i p \cdot (x-x')} \, e^{- \sigma\,  p^2 \, F(p^2) } \, . 
\ee
The function $F(p^2) = \Lambda_p/\Lambda_{k_0}$, encoding the quantum behavior, is determined by the value of the dimensionful  cosmological constant seen at energy scale $\sqrt{p^2}$ relative to its infrared (IR) value.
The short-distance behavior found in the LQG-inspired calculation of \cite{Modesto:2008jz} is a special case of \eqref{PQEG} with $F(p^2) = p^2$.

In HL gravity \cite{Horava:2009uw}, the quantum effects are encoded in a modified dispersion relation
$\omega^2 = f(\vec{p}^2)$. Herein the dependence of the function $f$ on the spatial momenta $\vec{p}$ is parametrized by
\be\label{PHL}
f(\vec{p}^2) = \vec{p}^2 \, \frac{1 + B \vec{p}^2 + C \vec{p}^4}{1 + D \vec{p}^2} \, .
\ee
The parameters $B, C, D$ have been determined through a comparison to three-dimensional CDT data \cite{Benedetti:2009ge} and in the following we will
use the ``preferred values''  $B=-1.18, C=344.47, D=10.08$ found in \cite{Sotiriou:2011mu}. The
anisotropic dispersion relation then results in the expression
\be
P_{\textsc{hl}}(x,x',\sigma) = \int \frac{d^2p\, d\omega}{(2\pi)^3} \, e^{i {\vec p} \cdot ({\vec{x}-\vec{x}'})}e^{i \omega t} \, e^{-\sigma \, [\omega^2 + f(\vec{p}^2)]} \, . 
\ee
Here the Wick-rotated time coordinate $t$ of the manifold is not to be confused with the external diffusion time $\sigma$.
A rather generic consequence of having nontrivial functions $F(p^2)$ and $f(\vec{p}^2)$ is the lack of an \emph{a priori} guarantee of a positive semidefinite solution: in general, the Fourier transform of a non-Gaussian function of the form $e^{- \sigma p^{2+\delta}}$ is not positive semidefinite. For illustrative purposes, we exemplify this feature based on the expressions \eqref{PQEG} and \eqref{PHL}. Since the functions $F$ and $f$ depend on the absolute value of the Euclidean momentum only, the angular integrations in momentum space can be carried out by applying
\begin{equation}
\int d^dp f(p)e^{-i p\cdot x} = \frac{\left( 2 \pi\right)^{\frac{d}{2}}}{r^{\frac{d}{2}-1}}\int_0^{\infty}\!\!\!dp\, p^{\frac{d}{2}} J_{\frac{d}{2}-1}\left( p  r\right) f(p),
\end{equation}
where $J_\nu(x)$ is the Bessel function of the first kind and order $\nu$ and $r = \vert x\vert$. In this way, the probability densities reduce to the Hankel transform 
of the modified exponentials. Explicitly,
\be\label{PQEGi}
P_{\textsc{qeg}}(r, \sigma) = \tfrac{r}{(2 \pi r)^{\frac{d}{2}}} \int_0^\infty dp\, p^{d/2} J_{\frac{d}{2}-1}(pr) \, e^{- \sigma\,  p^2 \, F(p^2) },
\ee
while
\be\label{PHLi}
P_{\textsc{hl}}(t, r, \sigma) = \frac{e^{- \frac{t^2}{4\sigma}}}{(4 \pi \sigma)^{1/2}} \, \int_0^\infty \frac{k \,dk}{2\pi} \, J_0(kr) \, e^{- \sigma f(k^2)} \, .
\ee

\begin{figure}[!ht]
\includegraphics[width=0.8\linewidth]{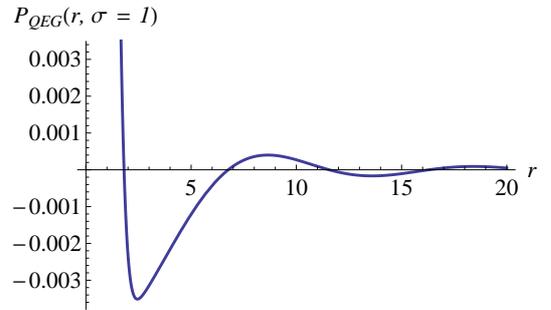}
\caption{\label{pqeg} The function $P_{\textsc{qeg}}(r, \sigma = 1)$, Eq.\ \eqref{PQEGi}, evaluated for a typical RG trajectory of quantum Einstein gravity in $d=3$.}
\end{figure}

\begin{figure}[!ht]
\includegraphics[width=0.8\linewidth]{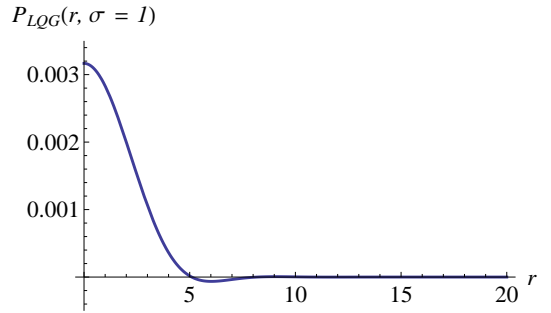}
\caption{\label{plqg} The function $P_{\textsc{lqg}}(r, \sigma = 1)$, Eq.\ \eqref{PQEGi}, evaluated for the ultraviolet (UV) limit in the LQG-inspired scenario of \cite{Modesto:2008jz}, by setting $d=4$ and $F(p^2) = p^2$.}
\end{figure}

\begin{figure}[!ht]
\includegraphics[width=0.8\linewidth]{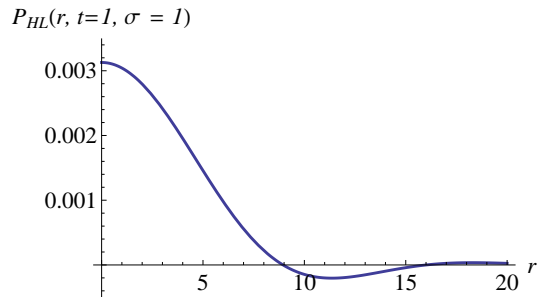}
\caption{\label{phl} The function $P_{\textsc{hl}}(r, t=1, \sigma = 1)$, Eq.\ \eqref{PHLi}, obtained from the anisotropic diffusion equation of HL gravity \cite{Sotiriou:2011mu}. The occurrence of negative values is a generic feature and does not rely on the special choice of parameters in \Eqref{PHL}.}
\end{figure}

The remaining integral in \Eqref{PHLi} can be evaluated numerically. For fixed values of $\sigma$ and $t$, snapshots of Eqs.\
\eqref{PQEGi} and \eqref{PHLi} are shown in Figs.\ \ref{pqeg}--\ref{phl}.
As illustrated by these plots, a common feature shared by all these modified diffusion equations is that the resulting probability densities for the probe particle are no longer positive semidefinite.
Thus, strictly speaking, the ``quantum-improved'' diffusion equations do not give rise to a well-defined diffusion process. Accordingly, one should proceed with great care when interpreting the resulting quantity $\ds$ as a spectral dimension. Moreover, at a conceptual level this feature constitutes a mismatch between the spectral properties derived within Monte Carlo studies and the improved diffusion equations. We take this as a motivation to investigate the possibility of ``quantum-improved'' diffusion equations 
which give rise to a manifestly positive-definite probability density in the next sections. As an additional bonus, the construction of these will also allow us to gain further insight into the nature of the diffusion process and the underlying stochastic interpretation in quantum gravity.

\section{Genuine diffusion on quantum spacetime}
\label{sect.3}
\noindent
The aim of this section is to construct diffusion equations which capture the quantum properties of spacetime while admitting solutions that are manifestly positive semidefinite. 
Essentially this can be achieved in two ways: either by considering diffusion in nonlinear time and/or modifying the quantum diffusion equation \eqref{absdiff} through a nontrivial source term. Both cases will be discussed below. For concreteness, we will focus on diffusion processes in asymptotic safety, but the results are also relevant for the UV limit of the LQG-inspired model of Ref.\ \cite{Modesto:2008jz}. The generalization to HL gravity will be discussed in Sec.~\ref{anisotropic}.
As a nontrivial finding, our novel diffusion equations recover all previous results on $\ds(\sigma)$.
\subsection{Diffusion in nonlinear time}\label{nonlineardiffusion}
\noindent
The crucial step in obtaining Eq.\ \eqref{absdiff} via an RG improvement is the identification of the cutoff scale $k$ with a suitable physical quantity. Such a procedure allows one to analyze the effects of quantum gravity in a variety of single-scale settings \cite{Reuter:2012xf}. Since we have seen that, generically, any quantum-gravity modification of the form $\nabla_x^2 \mapsto \langle \Delta \rangle$ will result in functions $P(x, x', \sigma)$
that have no interpretation as probability densities, we will explore an alternative RG-improvement scheme where the scale $k$ is related to the diffusion time $\sigma$. On general grounds, the identification $k = k(\sigma)$ should be such that large diffusion times correspond to the IR regime $k\rightarrow 0$ and short diffusion times to the UV regime $k \rightarrow \infty$. 

The RG-improved diffusion equation based on nonlinear time can be derived through a modification of 
the original RG-improvement scheme of Ref.\ \cite{Lauscher:2005qz}. For that purpose, we assume that
the scale-dependent solutions of the equations of motion based on the flowing action $\Gamma_k$
give rise to a power-law relation between the effective metrics at scale $k$ and the reference scale $k_0$, 
$\langle g^{\mu \nu}\rangle_k \propto k^\delta\, \langle g^{\mu \nu} \rangle_{k_0}$; see Appendix \ref{appA}.
Substituting this relation, Eq.\ \eqref{absdiff} becomes
\be\label{absdiff2}
\left(\partial_{\sigma} - k^\delta \langle g^{\mu\nu}   \nabla_\mu \nabla_\nu \rangle_{k_0} \right) P(x,x',\sigma)=0 \,, 
\ee
where $k$ is the RG scale, $k_0$ is the IR reference scale and $g^{\mu\nu}$ is the
fixed IR reference metric which we take to be the flat Euclidean metric. Since we want to encode the scaling effects in the diffusion time, we multiply the equation with $k^{-\delta}$ so that the diffusion operator becomes a standard second-order Laplacian,
\be\label{powerlawdiffmod}
\left(k^{-\delta} \frac{\partial}{\partial\sigma}-\nabla^2_x\right)P(x,x',\sigma)=0\,.
\ee
The relation between $k$ and $\sigma$ is then fixed on dimensional grounds.  Since $kx$ is dimensionless, Eq.\ \eqref{powerlawdiffmod} implies that, dimensionally, $\s\sim k^{-\delta-2}$ and suggests the scale identification
\be\label{cutoffid}
k = \sigma^{-\frac{1}{\delta + 2}},
\ee
where the proportionality constant has been absorbed into diffusion time $\sigma$.
By changing the diffusion time variable from $\sigma$ to $\sigma^\beta$ with
\be\label{betaeq}
 \beta = \frac{2}{\delta + 2},
\ee
the resulting equation can be cast into a diffusion equation in nonlinear time:
\be\label{powerlawdiff}
\left(\frac{\partial}{\partial\sigma^{\beta}}-\nabla^2_x\right)P(x,x',\sigma)=0\,.
\ee
The probability density resulting from this diffusion equation is given by
a Gaussian in $r = \vert x-x' \vert$:
\be\label{probdist2}
P(r, \sigma) = \frac{1}{(4 \pi \sigma^\beta)^{\frac{d}{2}}} \, e^{- \frac{r^2}{4 \sigma^\beta} } \, . 
\ee
Thus it is manifestly positive semidefinite. Moreover, the cutoff identification \eqref{cutoffid}
implies that $P[r, \sigma(k)] \propto k^d$ has the correct scaling behavior of a diffusion probability
in $d$ dimensions, giving further justification to the RG-improvement procedure.\footnote{This diffusion equation is also relevant to the case of loop quantum gravity, where the scaling of the area operator at small distances $l \ll l_0$ suggests to deduce a scaling of the metric in the form $g^{\mu \nu}(l) \sim l^{-2} g^{\mu \nu}(l_0)$ \cite{Modesto:2008jz}. Following the steps outlined above then allows one to arrive at \eqref{powerlawdiff} as a description of diffusion processes in the LQG-inspired model of \cite{Modesto:2008jz}.  However, caution should be exercised in drawing any conclusion regarding the spectral dimension in loop quantum gravity, since computations in a bottom-up approach (i.e., by placing a random walker in realistic LQG graphs) can lead to considerably more complicated results \cite{COT2}.}

 The spectral dimension resulting from \Eqref{probdist2}
is independent of $\sigma$ and reads
\be\label{dsde}
\ds = \frac{2d}{2+\delta}.
\ee
Notably, this expression is identical to the one derived from  RG improving
the Laplacian of the diffusion equation \cite{Reuter:2011ah}. Thus, the two RG-improved schemes give rise to the
same profile for the spectral dimension. This implies, in particular, that the matching between the spectral dimension
obtained from the diffusion in nonlinear time, \Eqref{powerlawdiff}, and the one measured within CDT is
identical to the one found in \cite{Reuter:2011ah}. Moreover, the positive semidefinite
function \eqref{probdist2} has the correct qualitative features
of a probability density measured within CDT, making a future comparison
between these more detailed spectral properties of the underlying 
quantum spacetimes meaningful.\footnote{While the probability density $P(x, x', \sigma)$ can certainly be measured 
on the quantum spacetimes underlying CDT, no such data are currently available.}

The averaged squared displacement of the test particle implied by \eqref{probdist2} is easily found to be
\be \label{r2pl}
\langle r^2 \rangle_\text{nonlinear time} = 2 \, d \, \sigma^\beta \,,
\ee
where angular brackets denote the expectation value with respect to the associated probability density function, $\langle f(x)\rangle=\int \rmd^d x\, P(x,x',\s)\,f(x)$. For $\beta = 1$, this corresponds to a standard Wiener process. In the case $\beta < 1$ 
the diffusion time passes slower, and slows down further throughout the diffusion process.
The diffusion process is subdiffusive, i.e., the average displacement of the test-particle is
less than in the diffusion on an ordinary $d$-dimensional manifold. 

There are actually two possible stochastic processes underlying Eq.\ \eqref{powerlawdiff}. One is scaled Brownian motion (SBM) \cite{LiM02,MeK04,Sok12}, i.e., a Wiener process which takes place in ``nonlinear time.'' The second is fractional Brownian motion (FBM) \cite{MeK04,Sok12,BaA,MaV},  which is a stochastic process with correlated increments, or, in other words, non-Markovian.  Let us note that, since both stochastic processes lead to the same diffusion equation, we cannot distinguish them at this point. They do however differ in one crucial property: SBM is actually Markovian, while FBM is not. Although we do currently not have enough data to rule out a non-Markovian stochastic process in either asymptotically safe gravity or the tentative continuum limit of CDTs, this property could be used to distinguish between the processes in the future.
Both processes and their properties (as well as a third twin, not likely to be relevant in the present context) are discussed in Appendix \ref{depr}. There we recall how diffusion processes are classified. This helps to develop a more intuitive understanding of the properties of quantum spacetime as probed by a diffusing particle. For instance, \Eqref{powerlawdiff} corresponds neither to trapped subdiffusion (where the test particle spends much time in bound states) nor to labyrinthine diffusion, where ``obstacles'' and ``holes'' are interposed by geometry and topology, as is in accordance with the expectation from asymptotically safe gravity. 
In particular, this does not correspond to
diffusion on a fractal in a mathematical sense.\footnote{Transport on fractals is realized by \emph{labyrinthine} diffusion \cite{Sok12}: subdiffusion is caused by the geometric and topological structure of the fractal and, so to speak, the test probe lives in a ``crowded'' environment and meets a number of obstacles and dead ends. Diffusion on a fractal is, in general, described by a diffusion equation with fractional differential operator $\p_\s^\beta$, see \Eqref{fractional_diff}, and an $x$-dependent diffusion coefficient plus friction (see \cite{MeN} and references therein).} The alternative interpretation is that subdiffusion occurs because of a viscoelastic effect: the test probe is dragged by the complex environment of which it is part. 
Intuitively, the subdiffusion can be understood as quantum fluctuations of spacetime slowing down the propagation of the particle as compared to a classical background. It is interesting to observe that this property is actually common between a variety of quantum gravity approaches, whereas no case is known in which quantum effects lead to superdiffusion. Let us emphasize again at this point that these findings do not allow to conclude that a physical particle propagating on the effective spacetime is subject to any similar effect.

\subsection{Diffusion employing fractional derivatives}\label{fractdiffusion}
\noindent
An interesting alternative to the cutoff identification \eqref{cutoffid}
relates the $k$ dependence of \Eqref{powerlawdiffmod} to the order of the
derivative operator. The resulting diffusion equation is
formulated in terms of fractional derivatives,
\begin{equation}\label{fractional_diff}
\left(\partial_\sigma^{\beta} - \nabla_x^2\right)P_\beta(x,x',\sigma)=0\,, 
\end{equation}
where $\partial_\sigma^{\beta} $ is the left Caputo derivative (see, e.g., \cite{cap1,Pod99,frc1} for more details)
\be\label{capu}
(\partial^{\beta} f)(\sigma) :=  \frac{1}{\Gamma(1-\beta)}\int_{0}^{\sigma} \frac{\rm d \sigma'}{(\sigma-\sigma')^{\beta}}\partial_{\sigma'}f(\sigma'),
\ee
and $0 < \beta \leq 1$. Formally, \Eqref{fractional_diff} is obtained by relating $k$ to the
Laplace momentum $s$ of the diffusion time $\sigma$. Concretely,
the Laplace transform of an ordinary derivative of a function $f(\sigma)$ is
given by  $\mathcal{L}[\p_\s f(\sigma)](s) = s \mathcal{L}[f(\sigma)] - f(0)$. Applying this to Eq.\ \eqref{powerlawdiffmod} and identifying
\be\label{cutoffid2}
k = s^{\frac{1}{\delta + 2}} \,,
\ee
then modifies the power of $s$ appearing in the right-hand side of the Laplace transform and yields
\be
s^{-\frac{\delta}{\delta+2}} \left\{s \mathcal{L}[P(x,x',\sigma)]-P(x,x',0)\right\} - \nabla_x^2 P(x,x',\sigma)=0.
\ee
 Comparing 
this  expression to the definition of the Laplace transform of the Caputo derivative,
\be\label{lapcap}
\mathcal{L}[\partial^{\beta}f(\sigma)](s)= s^{\beta} \mathcal{L}[f(\sigma)]- s^{\beta -1}f(0) \, ,
\ee
one is led to the RG-improved diffusion equation \Eqref{fractional_diff} with $\beta$ given by \Eqref{betaeq}.
Notably, the Laplace transform also fixes the type of fractional derivative appearing in
the diffusion equation, since, e.g., a Riemann-Liouville fractional derivative
would give rise to a different form of \Eqref{lapcap}. Also, using the Caputo fractional derivative ensures that the evolution can be formulated as a standard Cauchy problem.

Following \cite{mkrev,Zas3}, the general solution of \Eqref{lapcap} can be 
written in terms of a modified Laplace transform
\be\label{modlaptrafo}
P_\beta(r,\sigma) = \int_0^{\infty} {d}s\, A_\beta(s,\sigma) \, P_1(r, s) \, ,
\ee
where
\be\label{p1int}
 P_1(r, s) = \frac{e^{-\frac{r^2}{4s}}}{(4 \pi s)^{\frac{d}{2}}} \, 
\ee
is the solution of the standard heat equation \eqref{BM} in $d$ dimensions. Constructing $A_\beta(s,t)$ with the help of an inverse Laplace transform leads to a modified one-sided L\'evy distribution, which guarantees the existence
and positivity of $P_\beta(r, \sigma)$ as long as $P_1(r,s)$ is a proper probability density function \cite{mkrev}. 

Explicitly,  for an initial condition in the form of a delta distribution, the kernel $A_\beta(s,\sigma)$ can be represented in terms of a Fox function,
\be
A_\beta(s,\sigma) = \frac{1}{\beta \, s} H^{1,0}_{1,1}\left[ \frac{s^{1/\beta}}{\sigma} \left|
\begin{array}{c}
(1,1) \\
(1, 1/\beta) 
\end{array}
 \right. \right] \, .
\ee
Utilizing the series representation of $H^{1,0}_{1,1}$, the kernel can be written 
as an infinite sum:
\be\label{sumrep3}
A_\beta(s,\sigma) = \frac{1}{s} \sum_{n=0}^\infty \, \frac{(-1)^n}{\Gamma(1-\beta-\beta n) \Gamma(1+n)} \, \left( \frac{s}{\sigma^\beta} \right)^{1+n}.
\ee
Substituting this series into \Eqref{modlaptrafo} and switching to the integration variable $u = s/\sigma^\beta$, 
a straightforward computation shows that the spectral dimension $\ds(\sigma)$ resulting from \Eqref{fractional_diff} is again independent of $\sigma$ and given by \Eqref{dsde}. As in the case of nonlinear time, quantum effects lead to subdiffusion: 
\be 
\langle r^2 \rangle_{\rm fractional} = \frac{2d}{\Gamma(1+\beta)} \, \sigma^\beta \, .
\ee
Comparing this result with \Eqref{r2pl}, we observe that the difference in the coefficient can be removed by a rescaling of the diffusion time $\s$. Note that the two cases can be distinguished by considering either the probability density function itself or its higher moments, which carry nontrivial information because the process is not Gaussian.

We now focus on special cases in which the sum \Eqref{sumrep3} can be carried out explicitly. For 
$\beta = 1/2$, one obtains
\be\label{b12sum}
A_{1/2}(s,\sigma) = \frac{1}{\sqrt{\pi \sigma}} \, e^{-\frac{s^2}{4 \sigma}} \, ,
\ee
while $\beta = 1/3$ gives
\be\label{b13sum}
A_{1/3}(s,\sigma) = \frac{3^{2/3}}{\sigma^{1/3}} \, {\rm Ai}\left[\frac{s}{(3\sigma)^{1/3}} \right] \,,
\ee
where Ai is the Airy function. Substituting these results into \Eqref{modlaptrafo}, the corresponding probability densities are
\be\label{prob_frac}
P_{1/2}(r,\sigma)= \frac{1}{\sqrt{\pi \sigma}} \int_0^{\infty} {d}s\, \frac{e^{-\frac{s^2}{4 \sigma}}}{\left(4 \pi s \right)^{d/2}}e^{-\frac{r^2}{4s}} 
\ee
and
\be\label{prob_frac13}
P_{1/3}(r,\sigma)= \frac{3^{2/3}}{\sigma^{1/3}} \int_0^{\infty} {d}s\, \frac{{\rm Ai}\left[\frac{s}{(3\sigma)^{1/3}} \right]}{\left(4 \pi s \right)^{d/2}}e^{-\frac{r^2}{4s}} \, .
\ee
The first case applies to the UV fixed-point regime of asymptotic safety, whereas the second will play an important role for HL spacetimes. For $\beta = 1/2$, the integral in \Eqref{prob_frac} can be performed explicitly and gives rise to a linear combination of hypergeometric functions. While, for $d \ge 2$, the solution $P_{1/2}$ shows a pole at $r \rightarrow 0$, the integral of the probability for any finite yet arbitrarily small region is finite, thus having a meaningful interpretation as a probability density.

The probability density \eqref{prob_frac} actually admits an interesting interpretation in terms of an iterated Brownian motion (IBM) \cite{Bur92,OB1,BOS,OB2}.\footnote{This is a special case of a continuous-time random walk \cite{ScM}, where the length of the jump between one site and the next has finite variance and the waiting time between one jump and the next has a power-law distribution. The diffusion equation in this case is given by $(\p_\s^\beta-\nabla_x^2)P=0$,  where the fractional derivative is Caputo's. Subdiffusion occurs because the particle is trapped in bound states, where it spends far more time than in free motion; this type of subdiffusion is then called \emph{trapped}, or due to trapping.} This process is defined by a normal Brownian motion evaluated ``in random time.'' More precisely, given two independent Brownian motions $B_{1,2}$, $X_{\textsc{ibm}}(\sigma)= B_1 [\vert B_2(\s)\vert]$. For higher-dimensional IBM, the process $B_1$ is a $d$-dimensional vector while $B_2$ remains one-dimensional \cite{OB2}. 
A physical application of IBM is diffusion in a crack \cite{BuK}, which could help to develop an intuition for the effective properties of quantum spacetime. One could picture the effects of quantum gravity as turning a smooth background spacetime into an effective spacetime where diffusing particles are subjected to a randomized diffusion time, and diffuse as if in a ``crack'' formed by quantum fluctuations of the geometry.

\begin{figure}[!ht]
\centering
\includegraphics[width=8cm]{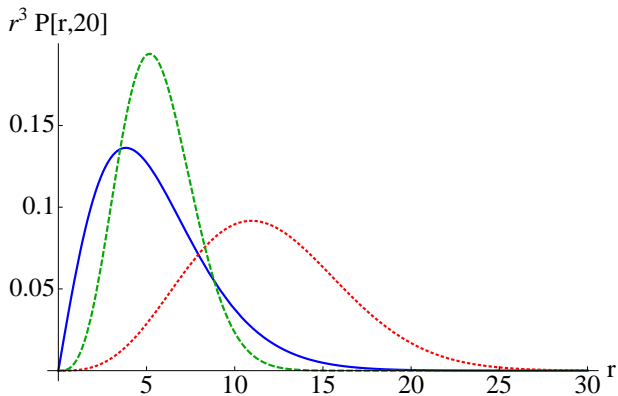}
\caption{\label{fig2} The probability density $P(r, \sigma=20)$ in $d=4$, weighted by $r^3$ for normal diffusion (red dotted line), power-law diffusion [\Eqref{powerlawdiff} with $\beta=1/2$, green dashed line] and fractional diffusion [\Eqref{fractional_diff} with $\beta=1/2$, thick blue line]. 
}
\end{figure}

At this stage, it is illustrative to compare the probability densities originating from normal diffusion \eqref{BM}, diffusion in nonlinear time \eqref{powerlawdiff}, and fractional diffusion \eqref{fractional_diff}. For the special case $d = 4,\, \beta = 1/2$ a snapshot taken at the diffusion time $\sigma = 20$ is shown in Fig.\ \ref{fig2}. This leads us to the following central observation: Both anomalous diffusion processes give rise to the same spectral dimension $\ds(\sigma)$. The two different physical processes can, however, be distinguished by their diffusion probability. Thus, we conclude that \emph{a matching spectral dimension does not imply that the quantum spacetime found in different approaches to quantum gravity is the same}. This  is a known fact from transport theory that should be appreciated in quantum gravity. As we discuss in Appendix \ref{depr}, a variety of different stochastic processes, including labyrinthine diffusion on fractals, continuous-time random walks, as well as fractional and scaled Brownian motion, can exhibit the same behavior of the mean square displacement and, hence, the same value for $\beta$. It is therefore currently unclear whether the common result $\ds=2$ in the UV regime of different quantum-gravity approaches is due to an \emph{accidental} degeneracy or, actually, arises due to a common physical origin.
Matching the spacetime picture emerging from different approaches to quantum gravity requires the analysis of more refined spectral quantities. The probability density of the diffusion process is one such quantity, providing a natural generalization of the spectral dimension.

\section{Diffusion on anisotropic spacetimes}
\label{anisotropic}

\noindent
The strategy of encoding the anomalous diffusion on an effective quantum spacetime by a nonlinear diffusion time or a fractional time derivative does not lend itself to an immediate generalization to the nonisotropic quantum spacetime underlying HL gravity, or any other anisotropic setting. In this case, it is an important observation \cite{OB1,BOS,AZ,DeB04,BMN1,Nan08} that the probability density \eqref{prob_frac} can also be obtained as a solution of an ordinary partial differential equation including higher-order spatial derivatives, once suitable source terms are included. In this section we generalize these ideas in order to obtain positive-definite solutions for the UV limit of HL gravity in three and four spacetime dimensions. The technical details of the computation can be found in Appendix \ref{App:B}. 

\subsection{HL spacetime with \texorpdfstring{$z=2,3$}{}}
\noindent
In \cite{Horava:2009if}, it was suggested that the diffusion equation describing the UV phase of HL gravity 
is given by
\be\label{Horava}
\left[\partial_{\sigma}- \p_t^2 - (\nabla_x^2)^z \right] \, P_{\textsc{hl}}(r,t,\sigma)=0 \, ,
\ee
where $z$ is the dynamical critical exponent capturing the anisotropic scaling in time and space
$t \rightarrow b t, \vec{x} \rightarrow b^z \vec{x}$. The value of $z$ is related to the 
dimension of spacetime $d= 1 + z$. The diffusion implied by \Eqref{Horava} is described by a
standard diffusion equation in (Euclidean) time, while the diffusion on the spatial slices
is anomalous. As explicitly shown in Sec.\ \ref{negprob}, the appearance of the higher order
spatial derivatives appearing in \Eqref{Horava} result in a $P_{\textsc{hl}}(r,t,\sigma)$ 
that is not positive semidefinite, and thus not a suitable probability density for 
a diffusion process.

Based on the results of the previous section, we construct a positive 
definite $P_{\textsc{hl}}(r,t,\sigma)$ as follows. In Appendix \ref{App:B}
it is shown that the positive definite probability densities
\eqref{prob_frac} and \eqref{prob_frac13} also solve the higher-order
partial differential equation
\be\label{IBMc}
\begin{split}
& \left[\partial_{\sigma}- (\nabla_x^2)^z \right]P_{1/z}(r,\sigma)= \cS_{1/z}(r,\s)\,.
\end{split}
\ee
The explicit form of the source terms $\cS_{1/z}$ is given in Eqs.\ 
\eqref{source12} and \eqref{source13}, respectively. Essentially, these correspond
to delta-function source terms which render the diffusion probability positive definite,
while leaving the spectral dimension unaffected.\footnote{The higher-derivative equation with $z=2$ without source terms has been
used to describe the diffusion of a test particle in the UV regime of asymptotically safe gravity \cite{Lauscher:2005qz, Reuter:2011ah}.} However, there is a new challenge in interpreting the source term, whose physical origin is presently unclear.

Comparing Eq.\ (\ref{IBMc}) and (\ref{Horava}), we see that we can match
the anisotropic diffusion equation by including a suitable time direction. Insisting
on the new $P_{\textsc{hl}}(r,t,\sigma)$ being positive definite requires a slight modification
of the source term, leading to
\be\label{IBM2}
\begin{split}
& \left[\partial_{\sigma}-  \partial_t^2 - (\nabla_x^2)^z \right]P_{\textsc{hl}}(t, r,\sigma)= P_1(t, \sigma) \, \cS_{1/z} (r, \sigma) \, .
\end{split}
\ee
The solution of \Eqref{IBM2} is the direct product of 
the Gaussian probability density in Euclidean time and the solution of anomalous diffusion in space with $\beta = 1/z$:
\be\label{PHL2}
P_{\textsc{hl}}(t, r, \sigma) = P_1(t, \sigma) \, P_{1/z}(r, \sigma) \, . 
\ee
As a product of two manifestly positive semidefinite probability densities, $P_{\textsc{hl}}$
is again a positive semidefinite quantity. For $z=2$, the spatial part of the probability density solves the fractional diffusion equation \Eqref{fractional_diff} and the temporal part follows a standard Wiener process (see Appendix \ref{App:B}). Thus, we can identify the underlying stochastic process as iterated Brownian motion in spatial directions and standard Brownian motion along the coordinate temporal direction.

The spectral dimension implied by the probability density \eqref{PHL2} is
\be\label{HLspec}
\ds = 1 + \frac{D}{z} \, , 
\ee
where $D$ denotes the dimension of the spatial slices. This is exactly the behavior expected for
the anisotropic spacetimes emerging within HL gravity \cite{Horava:2009if}. Our construction thereby reconciles the previous derivation with a positive semidefinite probability density. 

At this stage, it is interesting to pause for the following observation concerning the relation of HL gravity and the Monte Carlo simulations carried out within CDT. As shown by \Eqref{HLspec} the anomalous diffusion effects observed in HL have their origin in the spatial part of the diffusion equation, predicting that the diffusion on a spatial slice should lead to $\ds^{\textsc{hl}}|_{\rm spatial} = D/z$. For $3+1$ dimensional HL gravity where $D=3, z=3$ this implies $\ds^{\textsc{hl}}|_{\rm spatial} = 1$. This, however, seems to be in conflict with recent CDT measurements of the spectral dimension on spatial slices \cite{Gorlich:2011ga} which, for $D=3$, reported $\ds^{\textsc{cdt}}|_{\rm spatial} = 1.5$.  It would be interesting to study this apparent mismatch.

\subsection{Anomalous diffusion in space and time}
\noindent
Of course, the technique for generating a diffusion process that is anisotropic in space and time by tensoring the corresponding probability densities is not limited to the case where the diffusion in time is given by the probability density of a standard one-dimensional Brownian motion. More generally, one can also consider the case where diffusion is anomalous in both space and time with the anisotropy captured by two different diffusion coefficients $\beta_{\rm space}$ and $\beta_{\rm time}$. In this case, a positive semidefinite probability density could be obtained from
\be
P_{\rm anisotropy}(t,r,\sigma) = P_{\beta_{\rm time}}(t,\sigma) \times P_{\beta_{\rm space}}(r,\sigma) \, . 
\ee
Adapting the computation outlined in Appendix \ref{App:B}, it is straightforward to construct the corresponding anomalous diffusion equation satisfied by $ P_{\rm anisotropy}(t,r,\sigma)$. While this situation certainly goes beyond the framework of Ho\v{r}ava-Lifshitz gravity where $\beta_{\rm time} = 1$ by construction, this case could occur in other theories of quantum gravity. Since it is currently unclear if there is a suitable candidate theory of quantum gravity which realizes this situation, we refrain from a detailed analysis of this possibility at the present stage.

\section{Multiscale diffusion processes}
\noindent
So far, we have discussed anomalous diffusion processes with $\delta =\rm const$, implying that the spectral dimension of the effective quantum spacetime is the same at all length scales. In order to connect the quantum spacetime at sub-Planckian scales, where $\ds < d$, to the classical spacetime picture at long distances, where the spectral dimension should be equal to the topological one, $\ds = d$, $\delta$ should be promoted to a scale-dependent quantity. Generalizing $\delta\to\delta(\s)$ is not much different from what is sometimes done in transport theory, i.e., assuming a time-dependent anomalous exponent. This, for instance, was the proposal for multifractional Brownian motion \cite{LiM00,PLV,Lim01,LiM02} or other multiscale processes \cite{Zas3}.
For the remainder of this section we will focus on the multiscale geometries emerging within asymptotically safe gravity, \cite{Reuter:2011ah,Rechenberger:2012pm,Reuter:2012xf}. 

\begin{figure}[!ht]
\centering
\includegraphics[width=8cm]{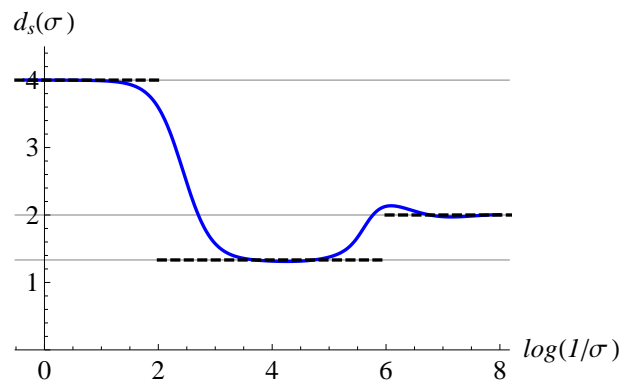}
\caption{\label{multiscale} Four-dimensional profiles of the scale-dependent spectral dimension $\ds(\sigma)$ obtained from \Eqref{dsde} with scale-dependent $\delta(k)$ (blue, solid curve) and the multiscale power-law diffusion equation \eqref{mseq} with $\sigma_1 = 0, \beta_1 = 1/2$, $\sigma_2 = 10^{-6}, \beta_2 = 1/3$, and $\sigma_3 = 10^{-2}, \beta_3 =1$ (black, dashed lines) Note that there is no dependence on the constants $c_i$ from \Eqref{mseq}.}
\end{figure}

In \cite{Reuter:2011ah}, $\delta(k)$ has been determined as a function of the scale-dependent Newton's constant and cosmological constant. This function, which can be evaluated numerically for a given RG trajectory and shows the three regimes $\delta =0, 4, 2$ as discussed above, is actually a prediction from the underlying fundamental theory, independent of the precise nature of the fictitious diffusion process used to sample the properties of spacetime. Thus, the scale dependence of $\delta$ is actually \emph{the same} for all diffusion processes considered in Sec.\ \ref{sect.3}. This implies, in particular, that diffusion in nonlinear time and with a fractional diffusion time gives rise to the same $\ds$ profile, which also agrees with the one constructed in \cite{Reuter:2011ah}. For a typical RG trajectory which runs towards positive values of the cosmological constant in the IR, the resulting $\ds(\sigma)$ is shown as the (blue) solid curve in Fig.\ \ref{multiscale}. The central feature of this curve is its distinguished plateaux where $\ds(\sigma)$ is approximately constant. 
From microscopic to macroscopic length scales, these are situated at $\ds = d/2$ (UV regime), $\ds = 2d/(2+d)$ (semiclassical regime), and $\ds = d$ (classical regime), and can be linked to universal features of the gravitational renormalization group flow. The plateaux are connected by short transition regimes.  The solution \eqref{probdist2} is valid except for these short regimes, i.e.,  in the plateaux $\delta \approx {\rm const}$ such that the power-law identification holds. 

Exploiting the fact that there is presumably no physical meaning attached to the transition regimes \cite{Reuter:2011ah},  this plateau structure can be mimicked in different ways. One is the simple multiscale power-law diffusion equation
\be\label{mseq}
\!\left[\sum_i \!c_i \theta (\sigma_{i+1}- \sigma) \theta(\sigma- \sigma_{i}) \frac{\partial}{\partial \sigma^{\beta_i}} - \nabla_x^2\right]\!\!P(r,\sigma)=0,
\ee
where $\theta$ are Heaviside step functions, $\sigma_i$ denote the transition scales between the distinct regimes, and the $c_i$ are constants of appropriate dimensionality which are fixed by the continuity and normalization of $P(r,\sigma)$. The diffusion-time operator could thereby be either the one for nonlinear time or fractional diffusion. In both cases the solution is given by a sum with step functions which glue together probability densities \eqref{probdist2} or \eqref{modlaptrafo} for different $\beta_i$. A typical example tailored to the plateau structure obtained from asymptotic safety is given by the black dashed curve in Fig.\ \ref{multiscale}.
Smooth profiles of multiscale geometries can be found in \cite{frc4,frc7}.
To find a smooth \emph{Ansatz}, the simplest way is to replace $\s^\beta$ in \Eqref{powerlawdiff} by a phenomenological profile $\ell^2(\s)$, so that
\be
\ds=d\frac{\p\ln \ell^2(\s)}{\p\ln\s}\,.
\ee
The function $\ell$ contains the hierarchy of fixed scales determining the onset of the various regimes. 

\section{Outlook: spectral properties from first principles}
\label{sect.6}
Up to this point, the discussion of the spectral properties of the quantum spacetime was based on the quantum-improvement of a classical diffusion equation. This constitutes a valid first step in capturing structural aspects of the quantum spacetime. Ultimately, one would like to derive the probability density $P(r, \sigma)$ directly from the underlying microscopic theory.

We expect that, within asymptotic safety, such a computation can be carried out by suitably adapting the framework of quantum diffusion in a stochastic medium; see \cite{Akkermans} for a pedagogical introduction. This framework relates the conditional probability $P(\vec{x}, \vec{x}', t)$ for a quantum-mechanical particle to propagate from state $\vec{x}$ to $\vec{x}'$ in time $t$ to the particle's Green's function:
\be\label{Qdiff}
\begin{split}
P(\vec{x}, \vec{x}', t) 
= & A^2 \int \frac{d\epsilon}{2\pi} \frac{d\omega}{2\pi}  \langle\!\langle G_{\epsilon}^{\rm R}(x,x') G_{\epsilon-\omega}^{\rm A}(x',x)\rangle\!\rangle \, \\
& \times e^{- \left[ (\epsilon + \omega/2 - \epsilon_0)^2/4\sigma_\epsilon^2 + (\epsilon - \omega/2 - \epsilon_0)^2/4\sigma_\epsilon^2  \right]} e^{i\omega t} \, . 
\end{split}
\ee
Here $G^{\rm R/A}_{\epsilon}(x,x')$ denotes the retarded/advanced Green's function with energy $\epsilon$, $\sigma_\epsilon$ is the width of the wave packet 
in the initial state and the double angular brackets denote an averaging over disorder. Assuming that the averaged Green's functions depend only slightly on $\epsilon$ and that $\omega$ is small
compared to $\sigma_\epsilon$, \Eqref{Qdiff} can be approximated by the Fourier transformed probability density,
\be\label{fpc}
P(x,x',\omega)= \frac{1}{2 \pi \rho_0} \langle\!\langle G_{\epsilon}^{\rm R}(x,x') G_{\epsilon-\omega}^{\rm A}(x',x)\rangle\!\rangle \,.
\ee
In practical computations, the average product of the two Green's functions is approximated by the product of the averaged Green's functions,
\be
\langle\!\langle G_{\epsilon}^{\rm R}(x,x') G_{\epsilon-\omega}^{\rm A}(x',x)\rangle\!\rangle \approx \langle\!\langle G_{\epsilon}^{\rm R}(x,x') \rangle\!\rangle \langle\!\langle G_{\epsilon-\omega}^{\rm A}(x',x)\rangle\!\rangle \, .
\ee
The Drude-Boltzmann approximation describes the probability of the particle to arrive at $x'$ without any collisions. 

Moving from quantum mechanics to asymptotic safety, it is natural to promote the diffusing particle to a scalar field living on the quantum spacetime. The Drude-Boltzmann approximation constitutes a natural analogue of the probe approximation
where the diffusing particle propagates on a fixed background.
Evaluating \Eqref{Qdiff} in the collisionless limit then requires computing the Green's function for this field in a situation
where fluctuations of spacetime are integrated out.  We expect that such a computation can be carried out along the lines of Ref.~\cite{Christiansen:2012rx}.
This should permit a first-principle evaluation of the probability density without the need of an \emph{a priori} postulate of a diffusion equation, since we expect the microscopic theory to determine the type of disorder, associated with the quantum fluctuations of spacetime. We hope to come back
to this point in the future.

\section{Conclusion}
In this work, we have used the diffusion of a fictitious probe particle to characterize the spectral properties of  quantum spacetimes. 
We have demonstrated that many quantum-improved diffusion equations proposed in the literature actually yield solutions that are not positive semidefinite. This deficit is traced back to the appearance of higher-derivative operators occurring in the ``quantum-improved'' procedure. Depending on the theory under consideration,
these operators emerge from an RG improvement of the classical diffusion kernel (asymptotic safety), the nontrivial scaling of the area operator in LQG-inspired scenarios,
or higher-order spatial derivatives capturing an anisotropic scaling of space and time (Ho\v{r}ava-Lifshitz gravity).
Based on this observation, we have proposed new classes of quantum-improved diffusion equations, relevant for all theories above, which can accommodate the quantum properties of spacetime. Their main virtue is to derive the profile of the (scale-dependent) spectral dimension $\ds(\sigma)$ obtained in earlier calculations while at the same time yielding a positive semidefinite probability density. These results place the derivation of $\ds(\sigma)$ on solid grounds while retaining all earlier conclusions \cite{Lauscher:2005qz, Reuter:2011ah,Rechenberger:2012pm,Modesto:2008jz,Horava:2009if,Sotiriou:2011mu}. This also allows us to investigate the probability density as a probe of quantum spacetime in more detail.
Moreover, we explicitly identify the underlying stochastic processes of two candidate equations as fractional, scaled, or iterated Brownian motion. Although the twin problem (i.e., the nonuniqueness of the underlying stochastic process to one of our proposals for the diffusion equation) is not solved at this stage, the physical insights entailed by the anomalous diffusion processes could constitute a first step in developing an intuitive picture of the underlying quantum spacetime.

As we have shown by the explicit construction of Eqs.\ \eqref{powerlawdiff} and \eqref{fractional_diff}, distinct diffusion equations with different underlying stochastic processes  (as well as different stochastic processes sharing the same diffusion equation) can lead to nearly identical profiles $\ds(\sigma)$. As discussed in \cite{frc4,fra7,frc7}, multifractional spacetimes can reproduce the same profiles, too, even in versions violating ordinary Lorentz symmetries. Thus, the spectral dimension constitutes only a very rough characterization of the quantum spacetime. A more refined picture has to go beyond the computation of $\ds(\sigma)$. In this work, we have taken the first step in this direction by constructing candidate probability densities for a particle probing the effective quantum spacetime of asymptotically safe gravity and Ho\v{r}ava-Lifshitz gravity. We expect that these probability densities will, at some point, also be accessible within other approaches to quantum gravity, foremost the Monte Carlo simulations underlying CDT and EDT, thereby offering a novel way to compare the spectral features of the resulting effective spacetimes. Our proposal complements and goes beyond the studies in \cite{Reuter:2011ah,Sotiriou:2011mu}, where the comparison was based on $\ds$ alone.

We close with the following cautious remark. Studying the quantum structure of spacetime based on quantum improving an \emph{ab initio} diffusion equation does not yield an unambiguous description of the dynamics of the probe particle. As we have shown explicitly, different improvement schemes lead to different probability densities. Therefore, ultimately a direct investigation of the probability density functions derived from first principles will be necessary. Based on this data, it can then be checked which effective diffusion processes could actually be realized by a given candidate theory for quantum gravity.\\ \bigskip

\section*{Acknowledgments} 

We thank Rafael Sorkin for critical questions that triggered this work, Andrzej G\"orlich, Petr Ho\v{r}ava, Samo Jordan, Jack Laiho, Renate Loll and Martin Reuter for helpful discussions and Ralf Metzler and Igor Sokolov for useful correspondence. The work of G.C.\ is under a Ram\'on y Cajal tenure-track contract. Research at Perimeter Institute is supported by the Government of Canada through Industry Canada and by the Province of Ontario through the Ministry of Research and Innovation. The research of F.S.\ is supported by the Deutsche Forschungsgemeinschaft (DFG)
within the Emmy-Noether program (Grant No.\ SA/1975 1-1).

\begin{appendix}
\section{ASYMPTOTICALLY SAFE GRAVITY-MATTER SYSTEMS}\label{appA}
 For asymptotically safe gravity \cite{Reuter:2012id}, the scale-dependent flowing action $\Gamma_k[g]$ implies a scale dependence of $\langle g_{\mu \nu} \rangle$ on the RG scale $k$. In the UV fixed-point regime\footnote{Here the notion of a scale-dependent metric makes sense, since, using $\langle g_{\mu \nu} \rangle_{k \rightarrow 0}$, which is the solution to the full quantum equations of motion, one can define the scale $k$ with respect to this unique, dynamically determined metric. This allows one to define a family of metrics $\langle g_{\mu \nu} \rangle_k$, which describe the effective background metric in an effective theory with cutoff scale $k$.}
\be\label{UVscaling}
\langle g_{\mu\nu}(x) \rangle_k \propto k^{-2} \qquad (k \rightarrow \infty) \, .
\ee
  %
%
The scaling relation \eqref{UVscaling} can be derived within simple truncations of the full effective action, but is actually an \emph{exact} consequence of asymptotic safety \cite{Lauscher:2005qz}, i.e., of scale invariance in the UV. The flowing action $\Gamma_k$ is given by an infinite sum of all operators $I_n$ compatible with the symmetries of the theory with their $k$-dependent coupling constants:
$ \Gamma_k = \sum_n \bar{g}_n(k) \, I_n [g_{\mu \nu}]$.
Passing to dimensionless couplings $g_n(k)= k^{-d_n}\bar{g}_n(k)$, with  the canonical dimensionality $d_n$ of the coupling, we can reabsorb the factor $k^{d_n}$ by noting that $I_n[k^2 g_{\mu \nu}] = k^{d_n}I_n[g_{\mu \nu}]$ under a scale transformation of the metric. Now we use the fact that in the asymptotic safety scenario the dimensionless couplings approach a fixed point in the ultraviolet, $g_n(k) \rightarrow g_{\ast}$: $
\Gamma_{k \rightarrow \infty}= \sum_n g_{\ast}\, I_n[k^2 g_{\mu \nu}]$. Thus, the $I_n$ only depend on the combination $k^2 g_{\mu \nu}$, so the solution $\langle g_{\mu \nu}\rangle_k$ to the effective equations of motion at large $k$ must obey a scaling property: If  $\langle g_{\mu \nu}\rangle_{k_0}$ is a solution at $k_0$, then it must be related to the solution  $\langle g_{\mu \nu}\rangle_k$ at $k$ by a simple rescaling,  $\langle g_{\mu \nu}\rangle_k \sim (k_0^2/k^2) \langle g_{\mu \nu}\rangle_{k_0}$. Crucially, this does not depend on the presence of a particular operator in the effective action. Thus, \Eqref{UVscaling} is a generic consequence of the theory becoming scale free in the UV.

Let us include matter degrees of freedom, and consider $I_n[g_{\mu \nu}, \Phi]$, where $\Phi$ stands for various matter fields (including gauge fields). Using the canonical form of the scale transformation of a matter field, the factor $k^{d_n}$ can be reabsorbed again. Thus, the argument given in \cite{Lauscher:2005qz} extends to the more realistic case of asymptotically safe gravity including arbitrary matter.

Let us spell out the argument in some more detail: The action of gravity coupled to matter can be schematically written as $\Gamma_k = \sum_{n,m} \bar{g}_{n,m}(k) I_n[g_{\mu \nu}]J_m[\Phi]$. Herein any operator can be written as a metric operator multiplied by a matter operator. The former can in this case have open indices, so the $I_n$ are a wider class of tensor structures than in the pure metric case.
Again, going over to dimensionless couplings, the effective action becomes $\Gamma_k = \sum_{n,m} g_{n,m}(k) k^{d_{g_{n,m}}}I_n[g_{\mu \nu}]J_m[\Phi]$. Here, the dimensionality of the coupling satisfies $d_{g_{n,m}}= - d_{I_n}- d_{J_m}$. Accordingly, under a scale transformation of the metric $g_{\mu \nu} \rightarrow k^2 g_{\mu \nu}$ and of the matter fields $\Phi \rightarrow k^{- d_{\Phi}} \Phi$, we get $J_m[\Phi] \rightarrow k^{-d_{J_m}} J_m[k^{-d_{\Phi}} \Phi]$ and $I_n[g_{\mu \nu}] \rightarrow k^{-d_{I_n}} I[k^2 g_{\mu \nu}]$. Therefore, we can again reabsorb the factor $k^{d_{g_{n,m}}}$ and get $\Gamma_k = \sum_{n,m} g_{n,m}(k)I_n[k^2 g_{\mu \nu}]J_m[k^{-d_{\Phi}}\Phi]$. In the fixed point regime, $g_{n,m}(k) \rightarrow g_{n,m \,\ast}$ and thus the action only depends on the combination $k^2 g_{\mu \nu}$. Again, we arrive at the conclusion that 
\be 
\langle g_{\mu \nu} \rangle_{k} = \frac{k_0^2}{k^2} \langle g_{\mu \nu} \rangle_{k_0}. \label{metricscaling}
\ee 
In addition, we conclude that, simultaneously, 
\be
\langle \Phi \rangle_k = \frac{k_0^{-d_{\Phi}}}{k^{-d_{\Phi}}} \langle \Phi \rangle_{k_0}.
\ee

In other words: If we have a regime where all dimensionful couplings scale according to their canonical dimensionality, then this regime can only be scale free, if the metric expectation value follows the inverse canonical Weyl rescaling, as in \Eqref{metricscaling}.

\section{DEGENERACY PROBLEMS IN DIFFUSION}\label{depr}

Here, we will discuss two degeneracy problems from  transport theory that are important in the context of quantum gravity:
\begin{itemize}
\item[(i)] Different stochastic processes can lead to the same result for the spectral dimension and the mean square displacement, as can be seen from cases (a), (b), (c), and (d) below.
\item[(ii)] Different stochastic processes can be described by the same diffusion equation and probability density function, as is the case for (c) and (d) which are relevant  for asympotically safe gravity. This is also known as the twin problem.
\end{itemize}
The fact that different models of quantum gravity sport the same spectral dimension in the UV has been enticing the community into looking for an explanation in terms of  common properties of the quantum spacetime. As we have discussed, the spectral dimension is a very rough probe, and the same value can emerge from widely different settings. However, the degeneracy of the numerical value of $\ds$ can be resolved when studying the diffusion process associated with these effective quantum spacetimes in greater detail. 

A variety of physical, biological, and biophysical systems display the phenomenon of subdiffusion (see \cite{Sok12} for references to the literature). In it, the mean squared displacement (or second moment, or variance) of the test particle is, in certain regimes, a power law in diffusion time:
\be\label{msd}
\langle X^2(\s)\rangle\propto \s^\beta\,,\qquad 0<\beta<1\,,
\ee 
where $\beta$ is a constant and $X(\s)$ (what we called so far the stochastic process) is the random variable describing the position of the particle at time $\s$. For normal diffusion (Brownian motion), $\beta=1$. 

Anomalous subdiffusion can be explained by transport models whose details should be determined by experiments on a case-by-case basis. The mean squared displacement \eqref{msd} can be reproduced by very different stochastic processes. We can recognize four main types of subdiffusion.
\begin{itemize}
\item[(a)] Transport on fractals is realized by \emph{labyrinthine} diffusion \cite{Sok12}: subdiffusion is caused by the geometric and topological structure of the fractal and, so to speak, the test probe lives in a crowded environment and meets a number of obstacles and dead ends. Diffusion on a fractal is, in general, described by a diffusion equation with fractional differential operator $\p_\s^\beta$ and an $x$-dependent diffusion coefficient, plus friction (see \cite{MeN} and references therein). 
\item[(b)] In continuous time random walk \cite{ScM}, the length of the jump between one site and the next has finite variance and the waiting time between one jump and the next has a power-law distribution. Subdiffusion occurs because the particle is trapped in bound states, where it spends far more time than in free motion; this type of subdiffusion is then called \emph{trapped}, or due to trapping. In the words of \cite{Sok12}, while the environment of fractal diffusion is a sort of rough landscape with a number of features, the environment of trap models is more akin to flat valleys surrounded by high ridges. The diffusion equation is of the form $(\p_\s-\p_\s^{1-\beta}\nabla_x^2)P=0$, where a constant diffusion coefficient is replaced by a fractional derivative in time \cite{mkrev}, or of the form $(\p_\s^\beta-\nabla_x^2)P=0$, where the first-order diffusion operator is replaced by a fractional Caputo derivative of order $\beta$. 
\item[(c)] \emph{Fractional Brownian motion} (FBM)
 \cite{BaA,MaV,WaT,LiM02} (see \cite{Sok12,Man82,MeK04} for reviews) and, more generally, diffusion associated with a generalized Langevin equation (GLE) \cite{WaT,LiM02,MeK04,KTH,Zwa01}. This class is characterized by a complex environment (e.g., a polymer network) of which the probe particle is just one part, interacting with the rest in a complicated way. These systems generically display viscoelastic behavior, and subdiffusion occurs because the probe is dragged by the environment. 

Fractional Brownian motion is defined by a GLE with a Weyl fractional integral,
\begin{eqnarray}
\p_\s X_\textsc{fbm}(\s) &=& \p_\s[{}_{-\infty}I^\g\eta](\s)\nonumber\\
&:=&\p_\s\int_{-\infty}^\s d\s'\, \frac{(\s-\s')^{\g-1}}{\Gamma(\g)}\,\eta(\s')\,,\label{gle2}
\end{eqnarray}
where $\eta$ is a Gaussian white noise.
This process starts at past infinity, is Gaussian and its two-point correlation function is $
\langle X_\textsc{fbm}(\s) X_\textsc{fbm}(\s')\rangle = d(|\s|^{2\g-1}+|\s'|^{2\g-1}-|\s-\s'|^{2\g-1})$, where $d$ is the topological dimension of space. For $\g>1/2$, the mean-squared displacement is
\be\label{msdfbm}
\langle X^2_\textsc{fbm}(\s)\rangle =2d |\s|^{2\g-1}\,;
\ee
the Hurst exponent $H$ in the literature is $H=\g-1/2$. As a model of subdiffusion ($1/2<\g<1$), FBM is said to be antipersistent \cite{Man82}, i.e.,  if the particle velocity is positive at one given step, it is more probable that it changes sign at the next step. When $1<\g<3/2$, FBM is persistent, and the sign of the velocity at two contiguous steps is most probably the same. On the other hand, in labyrinthine environments subdiffusion occurs because the probe bounces back against obstacles, and it inverts its motion.

FBM is the only stochastic process in ordinary space to be self-similar and characterized by stationary increments. Self-similarity means that time segments have the same behaviour at any time scale, after some rescaling:
\be\label{ssfbm}
X_\textsc{fbm}(\la\s)=\la^{\g-1/2} X_\textsc{fbm}(\s)\,.
\ee
Increments are stationary in the sense that their distribution depends only on the time interval:
\be
\langle [X_\textsc{fbm}(\s)-X_\textsc{fbm}(\s')]^2\rangle= \langle X^2_\textsc{fbm}(\s-\s')\rangle.
\ee
There exists also a version \cite{BaA,LiM02} with simpler GLE, such that $X_\textsc{rl-fbm}(\s):= [{}_0I^\g\eta](\s)$ features a fractional Riemann-Liouville integral. However, its increments are not stationary \cite{LiM02}.
\item[(d)] \emph{Scaled Brownian motion} (SBM) \cite{Sok12,MeK04,LiM02}, simply defined as a Brownian motion with power-law time, $X_\textsc{sbm}(\s) := B(\s^\beta)$. Although there are no known physical examples of it, scaled Brownian motion can be used as a fitting model of data with anomalous scaling, in cases where increments are nonstationary. It also has a remarkable interpretation, due to Sokolov \cite{Sok12}, as a Gaussian approximation for a ``cloud'' of continuous-time random walks.

In terms of the white noise $\eta(\s)$, the associated Langevin equation is
\be\label{gleb}
\p_\s X_\textsc{sbm}(\s) = \s^{\frac{\beta-1}2}\eta(\s)\,,
\ee
clearly different from Eq.\ \eqref{gle2}. SBM is self-similar, $X_\textsc{sbm}(\la\s)=\la^{\beta/2} X_\textsc{sbm}(\s)$, and for $\beta=2\g-1$ it reproduces both the FBM scaling law \eqref{ssfbm} and the anomalous mean-squared displacement \eqref{msdfbm},
\be\label{msdsbm}
\langle X^2_\textsc{sbm}(\s)\rangle= 2\,d\,\s^\beta\,.
\ee 
However, contrary to FBM, the increments of SBM are not stationary, since their distribution does not depend on the time interval only:
\begin{eqnarray}
\langle [X_\textsc{sbm}(\s)-X_\textsc{sbm}(\s')]^2\rangle&=&\langle [B(\s^\beta)-B({\s'}^\beta)]^2\rangle\nonumber\\
&=& \langle B^2(\s^\beta-{\s'}^\beta)\rangle\nonumber\\
&\neq& \langle X^2_\textsc{sbm}(\s-\s')\rangle\,.
\end{eqnarray}
This property is crucial to discriminate SBM from FBM experimentially, for instance via the first passage time distribution \cite{LiM02}. Another difference between the two processes is that SBM is Markovian, since the scale transformation $\s\to\s^\beta$ preserves time ordering for $\beta>0$ \cite{LiM02}. On the other hand, FBM is non-Markovian, due to the memory effect carried by the nonlocal fractional operator in its definition. 
\end{itemize}

The mean squared displacement alone is not sufficient to discriminate among these models and one has to look also at higher moments. These yield nontrivial information because the statistics of processes of type (a) and (b) are not governed by a Gaussian probability density function. In general, however, even the diffusion equation itself does not determine the stochastic process univocally, and there may exist processes (named twins) sharing not only the same variance scaling law, but also the same diffusion equation and probability distribution. An example is \Eqref{powerlawdiff}, which encodes \emph{both} an FBM and an SBM. In particular, from the diffusion equation alone it is not possible to tell whether a process is Markovian or not: its solution $P$ does not carry all the information of a stochastic process.

All known examples in quantum gravity are subdiffusive, which is why we concentrated on this case. Moreover, the spectral dimension $\ds$ is typically related to the exponent $\beta$ by $\beta=\ds/d_{\rm H}$, where $d_{\rm H}$ is the Hausdorff dimension of spacetime. Therefore, what was said above can be summarized in the two following statements. (A) Quantum gravity models with the same Hausdorff dimension can have the same spectral dimension even if the associated diffusion equations are quite different. (B) The twin problem may also be present, as in the case of \Eqref{powerlawdiff}.

Let us remark that the property of being Markovian cannot help us presently to distinguish which of the two processes is most probably related to asymptotically safe gravity, or to the description of the continuum limit in CDTs. In the first case, one would not expect that, e.g., the propagation of a physical particle on the quantum spacetime be non-Markovian, but, again, this cannot be used as an intuition for our setting. In CDTs, one should keep in mind that, although the discrete random walk on a CDT background is certainly Markovian, a non-Markovianness can still emerge in a nontrivial way in the continuum limit.

\section{DIFFUSION EQUATIONS WITH SOURCE TERMS}
\label{App:B}
In this Appendix we collect the technical details underlying the construction
of the partial differential equation (\ref{IBMc}). The construction
manifestly makes use of the integral representation of $P_{\beta}$, \Eqref{modlaptrafo},
for the two special cases $\beta = 1/2$ and $\beta = 1/3$.

The first step in the construction observes that the integral representation (\ref{modlaptrafo})
allows one to rewrite  derivatives with respect to $\sigma$ and $r$ of $P_{\beta}$ as derivatives with respect to $s$
appearing under the integral. Concretely, it is straightforward to verify that \Eqref{p1int}
satisfies
\be\label{sxrel}
(\nabla_x^2)^n \, P_{1}(r,s) = \partial_s^n P_{1}(r,s) \,,\qquad n = 1,2,\ldots\,,
\ee
while the kernels \eqref{b12sum} and \eqref{b13sum} solve the partial differential equations
\be
\partial_\sigma A_{1/2}(s, \sigma) = \partial_s^2 A_{1/2}(s, \sigma)
\ee
and
\be
\partial_\sigma A_{1/3}(s, \sigma) = - \partial_s^3 A_{1/3}(s, \sigma)\,,
\ee
respectively. Setting $n \equiv 1/\beta = 2,3$, we can use these relations to establish that
\be\label{ndereq}
\begin{split}
\p_\sigma P_{\beta} = & \int_0^\infty ds \left[\p_\sigma A_\beta(s, \sigma) \right] P_1(r,s) \\
= & \int_0^\infty ds \, (-1)^n \left[\p_s^n A_\beta(s, \sigma) \right] P_1(r,s) \\
= & \int_0^\infty ds \, A_\beta(s, \sigma) \left[\p_s^n P_1(r,s) \right] + \cS_{\beta} \\
= & \int_0^\infty ds \, A_\beta(s, \sigma) \left[(\nabla_x^2)^n P_1(r,s) \right] + \cS_{\beta} \\ 
= & \, (\nabla_x^2)^n P_{\beta}  + \cS_{\beta} \, .
\end{split}
\ee
The surface terms $\cS_{\beta}$ arising in the partial integration are given by
\be\label{surfacet}
\begin{split}
\cS_{\beta} = &  \left. \sum_{k=0}^{n-1} \, (-1)^{k+n} \, [ \p_s^{n-1-k} A_{1/n}(s, \sigma)] [\partial_s^k  P_1(r,s)]  \right|_0^\infty \\
= &  \left. \sum_{k=0}^{n-1} \, (-1)^{k+n} \, [\p_s^{n-1-k} A_{1/n}(s, \sigma)] [(\nabla_x^2)^k  P_1(r,s)]  \right|_0^\infty ,
\end{split}
\ee
where \Eqref{sxrel} has been used in the second step.

We now analyze the surface terms in more detail. The first observation is that, because of the exponential falloff of $A_\beta(s, \sigma)$ for $s \rightarrow \infty$, the upper boundary does
not give a contribution to $\cS_{\beta}$. On the lower boundary we actually have
\be
\lim_{s \rightarrow 0} \, P_1(r,s) = \delta^{(d)}(r) \, ,
\ee
which is the initial condition for the standard heat equation in $d$-dimensional Euclidean spacetime with diffusion time $s$. Exploiting that $P_{\beta}(r,\sigma)$ 
is actually subject to the same initial conditions, we can rewrite these terms as 
\be
\lim_{s \rightarrow 0} \, P_1(r,s) = P_{\beta}(r,0) \, .
\ee
Finally, we have to evaluate the $s \rightarrow 0$ limit of the kernels $A_{\beta}(s, \sigma)$ and its derivatives. For the case $\beta=1/2$, the surface terms
contain the zeroth and first derivative of \Eqref{b12sum}, given by
\be
\begin{split}
&\lim_{s \rightarrow 0} \, A_{1/2}(s, \sigma) =  \frac{1}{\sqrt{\pi \sigma}} \, , \\
&\lim_{s \rightarrow 0} \, \p_s A_{1/2}(s, \sigma) =  0 \, .
\end{split}
\ee
Substituting these results into \Eqref{surfacet} then yields the source terms for $\beta = 1/2$:
\be\label{source12}
\cS_{1/2} = \frac{1}{\sqrt{\pi \sigma}} \, \nabla_x^2 \, P_{1/2}(r,0) \,.
\ee
Here $P_{1/2}(r,0)$ denotes the $\sigma \rightarrow 0$ limit of \Eqref{prob_frac} and essentially
constitutes a delta-function source term. This source term agrees with the one found in \cite{AZ}.

Analogously, the case $\beta = 1/3$ requires evaluating 
\be
\begin{split}
&\lim_{s \rightarrow 0} \, A_{1/3}(s, \sigma) =  \frac{1}{\sigma^{1/3} \,  \Gamma(2/3)} \, , \\
&\lim_{s \rightarrow 0} \, \p_s A_{1/3}(s, \sigma) =  - \frac{1}{\sigma^{2/3} \, \Gamma(1/3)} \, , \\
&\lim_{s \rightarrow 0} \, \p_s^2 A_{1/3}(s, \sigma) =  0 \, .
\end{split}
\ee
The resulting surface term is then
\be\label{source13}
\begin{split}
\cS_{1/3} = & \, \frac{1}{\sigma^{1/3} \,  \Gamma(2/3)} \, (\nabla_x^2)^2 P_{1/3}(r,0) \\
& \, + \frac{1}{\sigma^{2/3} \, \Gamma(1/3)} \, \nabla_x^2 P_{1/3}(r,0)\,.
\end{split}
\ee
Substituting \Eqref{source12} or \Eqref{source13} into \Eqref{ndereq} finally yields
the  partial differential equation \eqref{IBMc} with $z=2,3$, completing our derivation.

\end{appendix}


\begin{thebibliography}{99}

\bibitem{Ambjorn:2005db} 
  J.~Ambj{\o}rn, J.~Jurkiewicz, and R.~Loll, \tia{Spectral dimension of the universe}
  \doin{10.1103/PhysRevLett.95.171301}{Phys.\ Rev.\ Lett.}{}{95}{171301}{2005} [\oarX{hep-th/0505113}].

\bibitem{Gorlich:2011ga} A.~G\"orlich, 
PhD thesis, Jagiellonian University, Krakow, 2010, \arX{1111.6938}.

\bibitem{Benedetti:2009ge} 
  D.~Benedetti and J.~Henson, \tia{Spectral geometry as a probe of quantum spacetime}
  \doin{10.1103/PhysRevD.80.124036}{Phys.\ Rev.}{D}{80}{124036}{2009} [\arX{0911.0401}].

\bibitem{Anderson:2011bj} 
  C.~Anderson, S.J.~Carlip, J.H.~Cooperman, P.~Ho\v{r}ava, R.K.~Kommu, and P.R.~Zulkowski,
  \tia{Quantizing Ho\v{r}ava-Lifshitz gravity via causal dynamical triangulations}
  \doin{10.1103/PhysRevD.85.049904}{Phys.\ Rev.}{D}{85}{044027}{2012} [\arX{1111.6634}].

\bibitem{Laiho:2011ya} 
  J.~Laiho and D.~Coumbe, \tia{Evidence for asymptotic safety from lattice quantum gravity}
  \doin{10.1103/PhysRevLett.107.161301}{Phys.\ Rev.\ Lett.}{}{107}{161301}{2011} [\arX{1104.5505}].

\bibitem{Modesto:2008jz} 
  L.~Modesto, \tia{Fractal structure of loop quantum gravity}
\doin{10.1088/0264-9381/26/24/242002}{Classical Quantum Gravity}{}{26}{242002}{2009} [\arX{0812.2214}].

\bibitem{Lauscher:2005qz} 
  O.~Lauscher and M.~Reuter,
  \tia{Fractal spacetime structure in asymptotically safe gravity} \doij{10.1088/1126-6708/2005/10/050}{J.\ High Energy Phys.}{10}{050}{2005} [\oarX{hep-th/0508202}].

\bibitem{Lauscher:2005xz} 
  O.~Lauscher and M.~Reuter,
  \tia{Asymptotic safety in quantum Einstein gravity: Nonperturbative renormalizability and fractal spacetime structure}
  in \emph{Quantum Gravity}, edited by B.~Fauser, J.~Tolksdorf, and E.~Zeidler (Birkh\"auser, Basel, Switzerland, 2007) 
  [\oarX{hep-th/0511260}].

\bibitem{Reuter:2011ah} 
  M.~Reuter and F.~Saueressig, \tia{Fractal space-times under the microscope: a renormalization group view on Monte Carlo data} \doij{10.1007/JHEP12(2011)012}{J.\ High Energy Phys.}{12}{012}{2011} [\arX{1110.5224}].

\bibitem{Rechenberger:2012pm} 
  S.~Rechenberger and F.~Saueressig, \tia{The $R^2$ phase-diagram of QEG and its spectral dimension}
  \doin{10.1103/PhysRevD.86.024018}{Phys.\ Rev.}{D}{86}{024018}{2012} [\arX{1206.0657}].

\bibitem{Mod11} L.\ Modesto, \tia{Superrenormalizable quantum gravity} \doin{10.1103/PhysRevD.86.044005}{Phys.\ Rev.}{D}{86}{044005}{2012} [\arX{1107.2403}].

\bibitem{Horava:2009if} 
  P.~Ho\v{r}ava, \tia{Spectral dimension of the universe in quantum gravity at a Lifshitz point} \doin{10.1103/PhysRevLett.102.161301}{Phys.\ Rev.\ Lett.}{}{102}{161301}{2009} [\arX{0902.3657}].

\bibitem{Sotiriou:2011mu} 
  T.P.~Sotiriou, M.~Visser, and S.~Weinfurtner,\tia{Spectral dimension as a probe of the ultraviolet continuum regime of causal dynamical triangulations} \doin{10.1103/PhysRevLett.107.131303}{Phys.\ Rev.\ Lett.}{}{107}{131303}{2011} [\arX{1105.5646}].

\bibitem{Car09} S.\ Carlip, \tia{Spontaneous dimensional reduction in short-distance quantum gravity?} 
  \doin{10.1063/1.3284402}{AIP Conf.\ Proc.}{}{1196}{72}{2009} [\arX{0909.3329}].
\bibitem{fra1}  G.\ Calcagni, \tia{Fractal universe and quantum gravity} \doin{10.1103/PhysRevLett.104.251301}{Phys.\ Rev.\ Lett.}{}{104}{251301}{2010} [\arX{0912.3142}].
\bibitem{Car10} S.\ Carlip, \tia{The small scale structure of spacetime} in {\it Foundations of Space and Time}, edited by G.\ Ellis, J.\ Murugan, and A.\ Weltman (Cambridge University Press, Cambridge, England, 2012) [\arX{1009.1136}].

\bibitem{Ambjorn:2012jv} 
  J.~Ambj{\o}rn, A.~G{\"o}rlich, J.~Jurkiewicz, and R.~Loll, \tia{Nonperturbative quantum gravity}
  \doin{10.1016/j.physrep.2012.03.007}{Phys.\ Rep.}{}{519}{127}{2012}
  [\arX{1203.3591}].

\bibitem{frc4}  G.\ Calcagni, \tia{Diffusion in multiscale spacetimes} \doin{10.1103/PhysRevE.87.012123}{Phys.\ Rev.}{E}{87}{012123}{2013} [\arX{1205.5046}].

\bibitem{Reuter:2012id} 
  M.~Reuter and F.~Saueressig, \tia{Quantum Einstein gravity}
  \doin{10.1088/1367-2630/14/5/055022}{New J.\ Phys.}{}{14}{055022}{2012}.

\bibitem{Reuter:1996cp} 
  M.~Reuter, \tia{Nonperturbative evolution equation for quantum gravity}
  \doin{10.1103/PhysRevD.57.971}{Phys.\ Rev.}{D}{57}{971}{1998} [\oarX{hep-th/9605030}].

\bibitem{Codello:2008vh} 
  A.~Codello, R.~Percacci, and C.~Rahmede,
  \tia{Investigating the ultraviolet properties of gravity with a Wilsonian renormalization group equation}
  \doin{10.1016/j.aop.2008.08.008}{Ann.\ Phys.\ (Amsterdam)}{}{324}{414}{2009} [\arX{0805.2909}].

\bibitem{Benedetti:2009rx} 
  D.~Benedetti, P.F.~Machado, and F.~Saueressig, \tia{Asymptotic safety in higher-derivative gravity}
  \doin{10.1142/S0217732309031521}{Mod.\ Phys.\ Lett.}{A}{24}{2233}{2009} [\arX{0901.2984}].

\bibitem{Akk12} E.\ Akkermans, \tia{Statistical mechanics and quantum fields on fractals} \arX{1210.6763}.
\bibitem{Akk2}  E.\ Akkermans, G.V.\ Dunne, and A.\ Teplyaev, \tia{Thermodynamics of photons on fractals} 
\doin{10.1103/PhysRevLett.105.230407}{Phys.\ Rev.\ Lett.}{}{105}{230407}{2010} [\arX{1010.1148}].

\bibitem{Groh:2010ta} 
  K.~Groh and F.~Saueressig,
  \tia{Ghost wave-function renormalization in asymptotically safe quantum gravity}
  \doin{10.1088/1751-8113/43/36/365403}{J.\ Phys.}{A}{43}{365403}{2010} [\arX{1001.5032}].

\bibitem{Eichhorn:2010tb} 
  A.~Eichhorn and H.~Gies,
  \tia{Ghost anomalous dimension in asymptotically safe quantum gravity}
  \doin{10.1103/PhysRevD.81.104010}{Phys.\ Rev.}{D}{81}{104010}{2010} [\arX{1001.5033}].

\bibitem{Vacca:2010mj} 
  G.P.~Vacca and O.~Zanusso,
  \tia{Asymptotic safety in Einstein gravity and scalar-fermion matter}
  \doin{10.1103/PhysRevLett.105.231601}{Phys.\ Rev.\ Lett.}{}{105}{231601}{2010} [\arX{1009.1735}].

\bibitem{Rosten:2011mf}  O.J.~Rosten, \tia{Relationships between exact RGs and some comments on asymptotic safety}
  \arX{1106.2544}.

\bibitem{Eichhorn:2012va} 
  A.~Eichhorn,
  \tia{Quantum-gravity-induced matter self-interactions in the asymptotic-safety scenario}
  \doin{10.1103/PhysRevD.86.105021}{Phys.\ Rev.}{D}{86}{105021}{2012}
  [\arX{1204.0965}].

\bibitem{Horava:2009uw}  P.~Ho\v{r}ava, \tia{Quantum gravity at a Lifshitz point}
 \doin{10.1103/PhysRevD.79.084008}{Phys.\ Rev.}{D}{79}{084008}{2009} [\arX{0901.3775}].

\bibitem{Reuter:2012xf} M.\ Reuter and F.\ Saueressig, \tia{Asymptotic safety, fractals, and cosmology} in \emph{Quantum Gravity and Quantum Cosmology}, edited by G.\ Calcagni, L.\ Papantonopoulos, G.\ Siopsis, and N.C.\ Tsamis  (Springer-Verlag, Berlin, 2013); \doin{10.1007/978-3-642-33036-0_8}{Lect.\ Notes Phys.}{}{863}{185}{2013} [\arX{1205.5431}].  

\bibitem{COT2}  G.\ Calcagni, D.\ Oriti, and J.\ Th\"urigen (work in progress).

\bibitem{LiM02} S.C.\ Lim and S.V.\ Muniandy, \tia{Self-similar Gaussian processes for modeling anomalous diffusion}
  \doin{10.1103/PhysRevE.66.021114}{Phys.\ Rev.}{E}{66}{021114}{2002}.

\bibitem{Sok12} I.M.\ Sokolov, \tia{Models of anomalous diffusion in crowded environments}
  \doin{10.1039/C2SM25701G}{Soft Matter}{}{8}{9043}{2012}.
	
\bibitem{MeK04} E.\ Metzler and J.\ Klafter, \tia{The restaurant at the end of the random walk: recent developments in the description of anomalous transport by fractional dynamics}
  \doin{10.1088/0305-4470/37/31/R01}{J.\ Phys.}{A}{37}{R161}{2004}.
	
\bibitem{BaA}   J.A.\ Barnes and D.W.\ Allan, \tia{A statistical model of flicker noise}
  \doin{10.1109/PROC.1966.4630}{Proc.\ IEEE}{}{54}{176}{1966}.  
\bibitem{MaV}   B.B.\ Mandelbrot and J.W.\ Van Ness, \tia{Fractional Brownian motions, fractional noises and applications} \doin{10.1137/1010093}{SIAM Rev.}{}{10}{422}{1968}.


\bibitem{MeN}   R.\ Metzler and T.F.\ Nonnenmacher, \tia{Fractional diffusion: exact representations of spectral functions} \doin{10.1088/0305-4470/30/4/011}{J.\ Phys.}{A}{30}{1089}{1997}.


\bibitem{cap1}  M.\ Caputo, \tia{Linear model of dissipation whose $Q$ is almost frequency independent-II} \doin{10.1111/j.1365-246X.1967.tb02303.x}{Geophys.\ J.\ R.\ Astron.\ Soc.}{}{13}{529}{1967}.
\bibitem{Pod99} I.~Podlubny, \emph{Fractional Differential Equations} (Academic Press, San Diego, 1999).
\bibitem{frc1}  G.\ Calcagni, \tia{Geometry of fractional spaces} \ndoin{http://intlpress.com/site/pub/pages/journals/items/atmp/content/vols/0016/0002/00024226/index.html}{Adv.\ Theor.\ Math.\ Phys.}{}{16}{549}{2012} [\arX{1106.5787}].

\bibitem{mkrev} 
  R.~Metzler and J.\ Klafter, \tia{The random walk's guide to anomalous diffusion: a fractional dynamics approach} \doin{10.1016/S0370-1573(00)00070-3}{Phys.\ Rep.}{}{339}{1}{2000}.
\bibitem{Zas3}  G.M.~Zaslavsky, \tia{Chaos, fractional kinetics, and anomalous transport}
 \doin{10.1016/S0370-1573(02)00331-9}{Phys.\ Rep.}{}{371}{461}{2002}.

%
\bibitem{Bur92} K.\ Burdzy, \tia{Some path properties of iterated Brownian motion} in {\it Seminar on Stochastic Processes, 1992}, edited by E.\ \c{C}inlar, K.L.\ Chung, and M.J.\ Sharpe (Birkh\"auser, Boston, 1993). 
\bibitem{OB1}   E.\ Orsingher and L.\ Beghin, \tia{Time-fractional telegraph equations and telegraph processes with Brownian time}
  \doin{10.1007/s00440-003-0309-8}{Probab.\ Theory Relat.\ Fields}{}{128}{141}{2004}.
\bibitem{BOS}   L.\ Beghin, E.\ Orsingher, and L.\ Sakhno, \tia{Equations of mathematical physics and compositions of Brownian and Cauchy processes}
  \doin{10.1080/07362994.2011.581071}{Stoch.\ Anal.\ Appl.}{}{29}{551}{2011} [\arX{1008.0928}].  
\bibitem{OB2}   E.\ Orsingher and L.\ Beghin, \tia{Fractional diffusion equations and processes with randomly varying time} \doin{10.1214/08-AOP401}{Ann.\ Probab.}{}{37}{206}{2009} [\arX{1102.4729}].
\bibitem{ScM}   H.\ Scher and E.W.\ Montroll, \tia{Anomalous transit-time dispersion in amorphous solids} \doin{10.1103/PhysRevB.12.2455}{Phys.\ Rev.}{B}{12}{2455}{1975}.
\bibitem{BuK}   K.\ Burdzy and D.\ Khoshnevisan, \tia{Brownian motion in a Brownian crack}
  \doin{10.1214/aoap/1028903448}{Ann.\ Appl.\ Probab.}{}{8}{708}{1998}.

\bibitem{AZ}    H.\ Allouba and W.\ Zheng, \tia{Brownian-time processes: the PDE connection and the half-derivative generator}
  \doin{10.1214/aop/1015345772}{Ann.\ Probab.}{}{29}{1780}{2001} [\arX{1005.3801}].
\bibitem{DeB04} R.D.\ DeBlassie, \tia{Iterated Brownian motion in an open set}
  \doin{10.1214/105051604000000404}{Ann.\ Appl.\ Probab.}{}{14}{1529}{2004}.  
\bibitem{BMN1}  B.\ Baeumer, M.M.\ Meerschaert, and E.\ Nane, \tia{Brownian subordinators and fractional Cauchy problems}
  \doin{10.1090/S0002-9947-09-04678-9}{Trans.\ Am.\ Math.\ Soc.}{}{361}{3915}{2009} [\arX{0705.0168}].
\bibitem{Nan08} E.\ Nane, \tia{Stochastic solutions of a class of higher order Cauchy problems in $\mathbb{R}^d$}
  \doin{10.1142/S021949371000298X}{Stochastics Dyn.}{}{10}{341}{2010} [\arX{0809.4824}].  


\bibitem{LiM00} S.C.\ Lim and S.V.\ Muniandy, \tia{On some possible generalizations of fractional Brownian motion}
  \doin{10.1016/S0375-9601(00)00034-7}{Phys.\ Lett.}{A}{266}{140}{2000}.
\bibitem{PLV}   R.-F.\ Peltier and J.\ L\'evy V\'ehel, \tia{Multifractional Brownian motion: definition and preliminary results} (\href{http://hal.inria.fr/inria-00074045/en}{unpublished}).
\bibitem{Lim01} S.C.\ Lim, \tia{Fractional Brownian motion and multifractional Brownian motion of Riemann--Liouville type} \doin{10.1088/0305-4470/34/7/306}{J.\ Phys.}{A}{34}{1301}{2001}.

%
\bibitem{frc7} G.\ Calcagni and G.\ Nardelli, \tia{Spectral dimension and diffusion in multi-scale spacetimes} \arX{1304.2709}.

\bibitem{Akkermans}
E.~Akkermans and G.~Montambaux, \emph{Mesoscopic Physics of Electron and Photons} (Cambridge University Press, Cambridge, England, 2011).

\bibitem{Christiansen:2012rx} 
  N.~Christiansen, D.F.~Litim, J.M.~Pawlowski, and A.~Rodigast,
  \tia{Fixed points and infrared completion of quantum gravity}
  \arX{1209.4038}.


\bibitem{fra7}  G.\ Calcagni, \tia{Multi-fractional spacetimes, asymptotic safety and Ho\v{r}ava--Lifshitz gravity} Int.\ J.\ Mod.\ Phys.\ A (in press) [\arX{1209.4376}].

\bibitem{WaT}   K.G.\ Wang and M.\ Tokuyama, \tia{Nonequilibrium statistical description of anomalous diffusion}
  \doin{10.1016/S0378-4371(98)00644-X}{Physica (Amsterdam)}{}{265A}{341}{1999}.

\bibitem{Man82} B.B.\ Mandelbrot, \emph{The Fractal Geometry of Nature} (Freeman, San Francisco, 1982), Chap.\ 27.

\bibitem{KTH}   R.\ Kubo, M.\ Toda, and N.\ Hashitsume, \emph{Statistical Physics II --Nonequilibrium Statistical Mechanics} (Springer-Verlag, Berlin, Germany, 1985).

\bibitem{Zwa01} R.\ Zwanzig, \emph{Nonequilibrium Statistical Mechanics} (Oxford University Press, Oxford, England, 2001).

\end{thebibliography}
\end{document}